\documentclass[%
reprint,
 amsmath,amssymb,
 aps,
]{revtex4-1}

\usepackage{graphicx}
\usepackage{dcolumn}
\usepackage{bm}
\usepackage{bm}
\usepackage{graphicx}
\usepackage{times}
\usepackage{soul}
\usepackage{color}
\usepackage{braket}
\usepackage{natbib}
\usepackage{floatrow}
\usepackage{physics}
\usepackage{comment}
\usepackage[normalem]{ulem}

\newcommand{\ba}{\begin{align}}
\newcommand{\ea}{\end{align}}
\newcommand{\B}{\mathbf}

\usepackage{color}
\usepackage{soul,xcolor}
\usepackage[normalem]{ulem}
\definecolor{black}{rgb}{0,0,0}
\definecolor{blue}{rgb}{0,0,1}
\definecolor{green}{rgb}{0,1,0}
\definecolor{red}{rgb}{1,0,0}
\definecolor{brown}{rgb}{0.4,0.2,0}
\definecolor{darkgreen}{rgb}{0,0.7,0}
\definecolor{darkblue}{rgb}{0.0,0.0,0.5}
\definecolor{red}{rgb}{1,0,0}
\definecolor{deepmagenta}{rgb}{0.8, 0.0, 0.8}

\newcommand{\rtext}[1]{{\color{red}#1}}

\newcommand{\hidetext}[1]{{{\color{blue}\sout{}}}}
\newcommand{\veck}{\mathbf k}

\newcommand{\vecp}{\mathbf p}
\newcommand{\vecq}{\mathbf q}

\usepackage[colorlinks,
            citecolor = blue,
						linkcolor = blue,
						urlcolor  = blue,
						anchorcolor = blue,
						breaklinks = true
						]{hyperref}

\def \ETH{Institute for Quantum Electronics, ETH Z\"urich, CH-8093 Z\"urich, Switzerland}
\def \MPI{Max Planck Institute of Quantum Optics, 85748 Garching, Germany}
\def \MCQST{Munich Center for Quantum Science and Technology, Schellingstrasse 4, 80799 M\"unich, Germany}

\begin{document}

\title{Exciton-polarons in two-dimensional semiconductors and the Tavis-Cummings model }
\author{\surname{Atac Imamoglu}}\email{imamoglu@phys.ethz.ch}\affiliation{\ETH}
\author{\surname{Ovidiu Cotlet}}\affiliation{\ETH}
\author{\surname{Richard Schmidt}}\affiliation{\MPI}\affiliation{\MCQST}

\preprint{APS/123-QED}

\begin{abstract}
The elementary optical excitations of a two-dimensional electron or hole system have been identified as exciton-Fermi-polarons. Nevertheless, the connection between the bound state of an exciton and an electron, termed trion, and exciton-polarons is subject of ongoing debate. Here, we use an analogy to the Tavis-Cummings model of quantum optics to show that an exciton-polaron can be understood as a hybrid quasiparticle -- a coherent superposition of a bare exciton in an unperturbed Fermi sea and a bright collective excitation of many trions.  The analogy is valid to the extent that the Chevy Ansatz provides a good description of dynamical screening of excitons and provided the Fermi energy is much smaller than the trion binding energy. We anticipate our results to bring new insight that could help to explain the striking differences between absorption and emission spectra of two-dimensional semiconductors.
\end{abstract}

\maketitle



Two-dimensional (2D) semiconductors~\cite{Heinz14-NatPhys} such as monolayers of transition metal dichalcogenides (TMD) have emerged as an exciting platform for investigating many-body physics and strong correlations~\cite{Xu16-NatRevMat,Urbaszek18-RMP}. Due to strong Coulomb interactions, the optical excitation spectra of neutral TMDs are dominated by tightly bound excitons. The small Bohr radius $a_B$ of TMD excitons leads to ultra-short radiative decay rates, in turn ensuring that in clean samples the exciton resonance is predominantly radiatively broadened~\cite{Back18-PRL,Scuri18-PRL}. Introduction of itinerant electrons (holes) into the monolayer dramatically modifies the nature of the optical spectra and leads to the emergence of a new absorption/reflection resonance near the energy of the three-body bound -- trion -- state of an exciton and an electron (hole) \cite{Urbaszek18-RMP}. It has been recently shown that the relevant elementary optical excitations in this limit are excitons that are dynamically dressed by Fermi sea electrons (holes), termed attractive or repulsive exciton-polarons~\cite{suris2003correlation,MSidler2017,Macdonald17}.

The connection between attractive exciton polaron (AP) and trion excitations has been the subject of ongoing debate. The oscillator strength for optical generation of a single isolated trion by diffraction limited resonant light is $f_T \sim f_x (k_{ph} a_T)^2$, where $f_x$ is the exciton oscillator strength, $a_T$ is the trion Bohr radius and $k_{ph} =E_T/(\hbar c)$ is the momentum of a photon resonant with the trion transition ($E_T$). While $f_t \ll f_x$, experiments show that the AP resonance acquires an oscillator strength that is comparable to $f_x$. The goal of this Letter is to shed new light on the relation between AP and trion excitations by making use of the Tavis-Cummings (TC) model of quantum optics~\cite{TC-model}.\\

\textbf{Tavis-Cummings model}

We start our analysis by recalling that the TC model describes an ensemble of $N_a$ two-level atoms with an energy splitting $\omega_{eg}$ between the ground ($\ket g$) and excited ($\ket e$) states coupled to a single cavity mode~\cite{TC-model} of frequency $\omega_c$. The interaction Hamiltonian of this system is given by
\begin{eqnarray}
   H_{int} &=& \sum_i g_c^i ( \sigma_{eg}^i a_c + h.c. ) ,
\end{eqnarray}
where $a_c^\dagger$ is the cavity creation operator and $\sigma_{eg}^i$ denotes the raising operator of the $i^{th}$ two-level atom. The cavity mode and atoms are coupled by the single-atom coupling rate $g_c^i$ which we, for simplicity, assume to be identical for all atoms in the following, $\forall i \; g_c^i = g_c$.

The lowest energy excitation spectrum of the TC model consists of $N_a - 1$ dark states at energy $\omega_{eg}$ and two polariton states that can be expressed as a superposition of bare cavity and atomic excitations. We refer to the lowest energy excited state as the lower polariton (LP) state, which can be expressed as
\begin{eqnarray}\label{LPAnsatz}
    \ket{\Phi_{LP}} &=& \left(\alpha a_c^\dagger  + \beta \sum_{i=1}^{N_a} \sigma_{eg}^i   \right) \ket{0}   .
\end{eqnarray}
Here the state $\ket{0}$ describes the vacuum of the cavity and all atoms in their ground state. We consider the case where the cavity frequency $\omega_c$ is blue detuned with respect to the atomic transition $\omega_{eg}$ by a detuning $\Delta = \omega_c - \omega_{eg}$.

In the limit when the detuning $\Delta$ is large compared to $g_c \sqrt{N_a}$, as well as the cavity ($\kappa_c$) and atomic ($\Gamma_{eg}$) decay rates, one finds
\begin{equation}
    \alpha =  g_c \sqrt{N_a}/\Delta.
\end{equation}
For this parameter range the LP state is a predominantly bright (symmetric) excitation of $N_a$ atoms, together with a small probability amplitude ($\alpha$) for a single cavity-photon excitation. The expression for $\alpha$ shows the well-known collective enhancement of cavity-atom coupling from $g_c$ to $g_c \sqrt{N_a}$. As a result of this enhanced coupling, the LP state is red-shifted in energy as compared to the $N_a - 1$ atomic dark states by an energy
\begin{equation}\label{ELP}
    E_{LP}=\alpha^2 \Delta = g_c^2  N_a / \Delta,
\end{equation}
provided that the cavity decay rate $\kappa_c \ll g_c \sqrt{N_a}$ (strong-coupling limit). It is important to emphasize that the LP resonance is insensitive to inhomogeneous broadening of atomic energy levels, provided that this broadening is smaller than $E_{LP}$.

Our simplified discussion of the TC model did not account for the spontaneous emission of the atoms: if the cavity-mode area is large compared to the square of the cavity-mode wavelength $\lambda_c$, the total spontaneous emission rate of the atoms is hardly modified. In the opposite limit where cavity-Purcell enhancement dominates the atomic decay, the atomic decay takes place predominantly through a two-step process where coherent excitation exchange between the atoms and the cavity is followed by cavity decay.

The TC model can be extended to a two-dimensional setting by assuming that the Fabry-Perot cavity consists of two parallel mirrors and the atoms are embedded in a 2D lattice with a period $d \ll \lambda_c$. In this case, the in-plane momentum of the polariton excitations constitutes a good quantum number. One may then define the collective atomic raising operator corresponding to an excitation with momentum $\veck$,
\begin{eqnarray}
    \sigma_{eg}(\veck)  &=&  \sum_j \sigma_{eg}^j e^{i \veck \mathbf{R}_j},
\end{eqnarray}
where the $\mathbf R_j$ denote the atomic lattice sites. The ansatz for the lower polariton branch of the two-dimensional TC model is then finally obtained by replacing $a_c$ by $a_c(\veck)$ and $\sum_j \sigma_{eg}^j$ with  $\sigma_{eg}(\veck)$ in Eq.~\eqref{LPAnsatz}.  \\

{\bf Exciton-polarons}

The Hamiltonian describing the interacting exciton-electron system in a TMD monolayer can be written as~\cite{MSidler2017,Macdonald17,CFey2020}

\begin{eqnarray}
\label{eqn:hamiltonian}
    H_{xe} &=& \sum_{\B k}\omega_\B k x^{\dagger}_{\B k} x_{\B k} + \sum_{\B k} \epsilon_\B k e^{\dagger}_{\B k} e_{\B k} \nonumber \\
    &+& \frac{v}{\cal A}\sum_{\B {k,k',q}}  x^{\dagger}_{\B {k+q}} x_{\B k} e^{\dagger}_{\B {k'-q}} e_{\B k'}.
\end{eqnarray}
Here,   $e_{\B k}$ and $x_{\B k}$ denote the annihilation operators of  electrons and excitons  with momentum $\B k$, respectively. The electronic dispersion is  $\epsilon_\B k=\B k^2/(2 m_e)$. The exciton dispersion $\omega_\B k=\B k^2/(2 m_x)$ is defined with respect to the exciton energy $E_x$.

The  contact coupling constant $v$ characterizes the short-range interaction between excitons and electrons and it is  related to the trion binding energy by the Lippmann-Schwinger equation
\begin{align}\label{LSE}
    v^{-1}=-\frac{1}{\cal A}\sum_{|\B k|<\Lambda} \frac{1}{E_T + \omega_\B k + \epsilon_\B k}.
\end{align}
Here $E_T = \hbar^2/(2 m a_T^2)$ denotes the trion binding energy, and $1/m = 1/m_x + 1/m_e$ is the reduced mass. As evident from Eq.~\eqref{LSE}, the interaction is regularized by a UV cutoff $\Lambda$ which can physically be related to  the inverse Bohr radius of the exciton. However, assuming that the exciton Bohr radius is the smallest length-scale in the problem, one may take the limit $\Lambda \to \infty$ at the end of the calculation.

It has been shown \cite{Giraud2008,Trefzger2012} that the eigenstates of the interacting polariton-electron system can be accurately described using the variational Chevy ansatz~\cite{chevy2006universal}
\begin{eqnarray}\label{APAnsatz}
    \ket{\Psi_{AP, \bf p}} &=& \left(\phi_\B p x^\dagger_\B p + \sum_{\B k \B q} \phi^\vecp_{ \B k \B q} x^\dagger_{\B p + \B q - \B k} e^\dagger_\B k e_\B q  \right) \ket{\Phi}  \\
    &=& \left(\phi_\B p x^\dagger_\B p + \sum_{\nu \B q} \eta^\vecp_{\nu  \B q} t^\dagger_{\nu \B p + \B q}  e_\B q  \right) \ket{\Phi} ,
\end{eqnarray}
which expands the wavefunction in excitations of the non-interacting ground state $\ket{\Phi}$ of the electron system in the TMD monolayer. The variational ground state $\ket{\Psi_{AP, \bf p}}$ is the so-called attractive polaron (AP) of momentum $\vecp$ which describes the exciton as a quasiparticle dressed by the attractive interactions with the Fermi sea of electrons.

In the second line of Eq.~\eqref{APAnsatz}, we have introduced the  creation operator $t^\dagger_{\nu\B l}$ that generates a composite trion state of center-of-mass momentum $\B l$ in an internal state $\nu$:
\begin{align}
      t^\dagger_{\nu \B l} = \sum_{\B k} \chi^\B l_{\nu  \B k} x^\dagger_{\B l - \B k} e^\dagger_{\B k} .
\end{align}
Here the states $\nu$ denote both bound trion as well as electron-exciton scattering states \cite{CFey2020}. Importantly, while the sum over $\nu$ in Eq.~\eqref{APAnsatz} runs over all these composite states, for excitations around the  AP resonance  the  bound trion  state is the most relevant one ($\nu = 0$). For a zero-momentum AP,  Eq.~\eqref{APAnsatz} can therefore be expressed as
\begin{eqnarray}\label{APAnsatzReduced}
    \ket{\Psi_{AP,\vecp=0}} &\approx&  \left(\phi_0 x^\dagger_\B 0 + \chi_0 \sum_{ \B q} \tilde\eta_{ \B q} t^\dagger_{ \B q}  e_\B q  \right) \ket{\Phi},
\end{eqnarray}
where $\phi_0 \equiv \phi_{\vecp =0}$, $\chi_0 \tilde \eta_\vecq \equiv \eta^{\vecp=0}_{\nu=0,\vecq}$. Eq.~\eqref{APAnsatzReduced} can be interpreted as describing a  quasiparticle where an exciton with momentum $\B p = 0$ is hybridized with a collective optical excitation of all electrons in the Fermi sea. As we will show below, for low electron densities, where the Fermi momentum $k_F$ satisfies $k_F^2 a_T^2 \ll 1$, an AP excitation has a small probability amplitude ($\phi_0$) for a bare exciton excitation.  \\


{\bf Correspondence of the TC and exciton-polaron models}

The form of the LP and AP wavefunctions given in Eq.~\eqref{LPAnsatz} and Eq.~\eqref{APAnsatzReduced} already hint at a one-to-one correspondence between the elementary excitations occurring in rather different experimental systems. The equivalence of these two models can be clarified by identifying the correspondence between the operators
\begin{center}
\begin{tabular}{ l c l }
$\sigma_{eg}^i$ & $\iff$ & $\tilde \eta_\vecq t^\dagger_\vecq e_\vecq$ \\[2.5mm]
 $a_c^\dagger$ & $\iff$  & $x_{\mathbf{0}}^\dagger$
\end{tabular}
\end{center}
and key parameters
\begin{center}
\begin{tabular}{ l c l }
$\omega_c$ & $\iff$  & $E_x$\\[2.5mm]
$\omega_{eg}$ & $\iff$  & $E_x-E_T$ \\[2.5mm]
$\Delta$ & $\iff$ & $E_T$ \\[2.5mm]
$N_a$ & $\iff$ & $N_e = \mathcal{A}n_e=\mathcal{A}k_F^2 /(4\pi)$ \\[2.5mm]
$E_{LP}$ & $\iff$ & $E^{AP}$\\[2.5mm]
\end{tabular}
\end{center}
where $n_e=k_F^2/4\pi$ denotes the electron density in a single valley, $\mathcal{A}$ is the area of the TMD monolayer, $E_F = \hbar^2k_F^2/2m_e$ is the Fermi energy, and $E^{AP} = E_x - E_T - E_{\vecp = 0}^{AP}$ is the energy difference between the AP and trion resonances.
The wave functions of the LP and AP are related by
\begin{center}
\begin{tabular}{ l c l }
 $\beta$& $\iff$ & $\chi_0$ \\[2.5mm]
 $\alpha$ & $\iff$  & $\phi_0$.
\end{tabular}
\end{center}.

From this correspondence between the two models one would expect $E^{AP}$ to satisfy an expression similar to the one for $E_{LP}$ in Eq.~\eqref{ELP}. In fact, without calculation, the correspondence would directly imply that $E^{AP} = \phi_0^2 E_T$.  In the following, we demonstrate that this is indeed the case by an explicit calculation.



To this end, we use the electron-exciton scattering T-matrix that accounts for effects of the finite electron density~\cite{suris2003correlation,schmidt2012fermi,cotlet2018transport},
\begin{align}
    T(\B p, \omega)^{-1} = v^{-1} - \frac{1}{\cal A}\sum_{|\B k|>k_F} \frac{1}{\omega-\epsilon_{\B k} -\omega_{\B p-\B k}},
\end{align}
where $\B p$ and $\omega$ denote the total momentum and energy of the exciton and the electron (we set $E_x = 0$ for convenience). The exciton self-energy is obtained from the T-matrix as:
\begin{align}
    \Sigma_{x}(\B p ,\omega) =\frac{1}{\cal A}\sum_{|\B q|<k_F} T(\B p + \B q, \omega+ \epsilon_\B q).
\end{align}
The quasi-particle weight $|\phi_\B p|^2$ in turn is given by
\begin{align}
    |\phi_ {\B p}|^2 = \left( 1 -  \frac{\partial}{\partial \omega} \left[\Sigma_{x}(\B p ,\omega) \right]_{\omega=E^{AP}_\B p} \right)^{-1} ,
    \label{wavefunction}
\end{align}
where $E^{AP}_\vecp$ denotes the energy of the AP at momentum $\vecp$, as determined by the solution of the Dyson equation
\begin{equation}\label{DSE}
  \left.\left[  \omega - \omega_\vecp - \Sigma_{x}(\vecp,\omega)\right]\right|_{\omega = E^{AP}_\vecp} =0.
\end{equation}

We now focus on zero momentum $\vecp=0$. To obtain an analytical expression for $\phi_{\B p=0}$, we consider the low electron density limit where $E_T \gg E_F$. As shown in App.~A in this limit the exciton self-energy can be approximated by
\begin{eqnarray}\label{SimpSig}
\Sigma_{x}(\omega)=\Sigma_{x}(\vecp=\mathbf 0,\omega) \simeq n_e T_{xe}(0,\omega)
\end{eqnarray},
where the two-body T-matrix is given by
\begin{eqnarray}
    T_{xe}(0, \omega) &=& \frac{2 \pi \hbar^2}{m} \frac{1}{ln[\frac{E_T}{\omega}] + i \pi}\label{T2B} \\
    &\simeq& \frac{2 \pi \hbar^2}{m} \frac{E_T}{E_T - \omega}\label{TxeSimp} ,
\end{eqnarray}
provided  $E_T - \omega \ll E_T$.
Using this approximate expression for $T_{xe}(0, \omega)$ evaluated at $\omega = E_{\vecp = 0}^{AP}$, we obtain
\begin{align}
    |\phi_ {\B p=0}|^2 \simeq \frac{(E^{AP})^2}{E_T} \frac{m}{n_e 2 \pi \hbar^2} .
    \label{wavefunction}
\end{align}

To express $\phi_ {\B p=0}$ in terms of the Fermi momentum $k_F$ and $a_T$, we use the fact that the AP resonance energy $E_{\vecp = 0}^{AP}$ is given by the lowest energy pole of the exciton propagator, i.e. the solution of Eq.~\eqref{DSE}. In the limit $|E_T-E^{AP}_{\B p=0}| \ll E_T$, we obtain
\begin{equation}
E^{AP} = E_x - E_T-E^{AP}_{\B p=0} = n_e \frac{2 \pi \hbar^2}{m}  =   \frac{m_e}{m} E_F \; \; .
\label{energyshift}
\end{equation}
Substituting for $E^{AP}$ in Eq.~\eqref{wavefunction}, we thus  arrive at the expression,
\begin{align}
    \phi_ {\B p=0}^2 = k_F^2 a_T^2 .
    \label{wavefunction2}
\end{align}
Eq.~\eqref{wavefunction2} shows that the AP resonance has a collectively enhanced oscillator strength $f_{AP} =   k_F^2 a_T^2 f_x$ . Finally, using Eq.~\eqref{wavefunction2} and Eq.~\eqref{energyshift}, we find $E^{AP} = \phi_0^2 E_T$, verifying the perfect correspondence between the LP and AP resonances of the two models.  \\

{\bf Discussion of limitations}

We emphasize that despite the remarkable correspondence between the LP and AP resonances, the analogy between exciton-polarons and the TC model breaks down for the repulsive polaron branch owing to the logarithmic energy dependence of the exciton-electron  T-matrix, and consequently of $\Sigma_{xe}(\B p ,\omega)$. In contrast, the photon self-energy in the TC model is given simply by $g_c^2 N_a /(E-\omega_{eg})$.

The approximation $\Sigma_{x}(\omega,\vecp=\mathbf 0) \simeq n_e T_{xe}(0,\omega)$ we used  is valid either in the limit of low electron density $n_e$ or if the electron mass were much larger than the exciton mass; this would be the case if the monolayer is embedded in a 2D cavity with a small cavity length where the elementary excitations are exciton-polaritons with a very light effective mass. In the absence of a cavity, however, the fermionic nature of the electrons leads to a broadening of the trion transition of order $E_F$, which is in turn comparable to the shift of the AP resonance $\phi_ {\B p=0}^2 E_T$. Consequently, and unlike in the ideal TC model, the trion-hole pairs that contribute to the AP resonance do not have identical energy. Nevertheless, within the Chevy description,  the AP resonance is insensitive to this broadening  even for finite $n_e$ and is broadened exclusively by radiative decay arising from its bare exciton character.

The analogy we developed uses the simplest Ansatz for describing correlated exciton-electron states. In particular, this Chevy Ansatz does not capture the screening of trions by the Fermi sea of electrons (for a discussion in the context of ultracold atoms see Ref.~\cite{Punk2009}); indeed, neglecting Coulomb repulsion between electrons, it has been shown theoretically that dynamically screened trions have lower energy than AP provided $k_F a_T \leq 0.1$~\cite{Parish2011,Parish2013,Kroiss,Vlietinck}. Since this is the regime we consider in this work, our results will not be quantitatively accurate.

Arguably, the most important difference between the exciton-electron system and the 2D TC model is the drastic reduction of coupling of high momentum collective atomic excitations to the corresponding cavity modes in the TC model. Since the effective cavity photon mass is orders of magnitude smaller than that describing the collective atomic excitations, only the bright symmetric atomic excitation couples appreciably to the cavity mode. In the limit of a 0D cavity, this description becomes exact and justifies referring to the $N_a - 1$ antisymmetric excitations with energy $\omega_{eg}$ as dark states. As we highlighted earlier, these excitations are dark towards coupling to the cavity mode but in the limit of a cavity mode area $A_c \gg \lambda_c^2$ they will decay by the single-atom spontaneous emission rate.

Due to the comparable effective masses of the exciton, electron and the trion, each trion-hole pair state with total momentum $\B p$ ($t_{\nu \B p + \B q}^\dagger  e_\B q$) hybridizes with the exciton mode $x_{\B p}$. Therefore, in the limit $k_F a_T \ll 1$, only polaron states couple to light. Moreover, this coupling is proportional to the quasi-particle weight $|\phi_{\B p}|^2$, i.e. it is exclusively due to the bare-exciton character of the polaron. This argument of course does not preclude possible observation of single trion decay in a nonequilibrium experiment such as photoluminescence \rtext{\cite{glazov2020optical}}; if we ignore the dynamical screening of trions, each single trion-hole pair state is a superposition of AP eigenstates. Even in the presence of a finite electron density, an optically generated electron-hole pair could form a trion with a single electron and subsequently decay by emitting a photon before the excitation is spread throughout the sample to form the  AP state at $\vecp \simeq 0$.

In summary, we have developed an analogy between the interacting exciton-electron problem in 2D materials and the TC model. Our work shows that the AP resonance can be described as a hybridization of collective trion-hole pair excitations with excitons, which in turn ensures their enhanced coupling to external light fields. We emphasize that a simplistic picture describing the total optical absorption strength as having contributions from all $N_e = A_L k_F^2/(4 \pi)$ electrons within the excitation spot with area $A_L$ yields a similar result as what we obtain using the polaron model. However, such a description would erroneously predict line broadening by $E_F$ and misses out on the energy shift of the AP resonance ($E^{AP}$) from the single trion energy.  Finally, the collective nature of the polaron excitation with minimal disturbance of each electron ensures that  polaron excitation constitutes an invaluable nondestructive spectroscopic tool for investigating strongly correlated states of electrons.\\

{\bf Acknowledgements}

This work was initiated during an informal meeting with M. Glazov, M. Semina, B. Urbaszek, X. Marie, T. Amand, A. Högele and M. Kroner. We particularly acknowledge many useful discussions with M. Glazov. R.~S. is supported by the Deutsche Forschungsgemeinschaft (DFG, German Research Foundation) under Germany's Excellence Strategy -- EXC-2111 -- 390814868.\\

\appendix

\section{Exciton self-energy at low doping}\label{AppTMatrix}

As shown in Ref.~\cite{Combescot2007}  the solution to the variational ansatz is equivalent to a non-selfconsistent resummation of ladder diagrams. Thus we may express our results by using the language of many-body field theory in terms of T-matrices and selfenergies. Generalizing Ref.~\cite{schmidt2012fermi} to the mass-imbalanced case one finds the expression for the exciton selfenergy (in this Appendix we work in units $\hbar =1$)
\begin{widetext}
\begin{eqnarray}
\Sigma_x(\vecp=\mathbf{0},\omega)=\int_{|\vecq|<k_F} \frac{d^2q}{(2\pi)^2}  T_{xe}\Big[\frac{1}{2}\Big(\omega+\epsilon_\vecq+\omega_\vecq-2\epsilon_\vecq^\text{Tot}-\epsilon_F^R+i 0^+-\sqrt{(\omega+\epsilon_\vecq-\omega_\vecq-\epsilon_F^R)^2 - 4 \epsilon_F^x \omega_\vecq- i 0^+}\Big)\Big],\nonumber\\\label{TStart}
\end{eqnarray}
\end{widetext}
where we define $\epsilon_\vecq^\text{Tot}=\vecq^2/(m_x+m_e) $, $\epsilon_F^x = k_F^2/(2m_x)$, and $\epsilon_F^R = k_F^2/(2m)$. Since the integral extends only up to $|\vecq|<k_F$ and the pole of the polaron will be in the vicinity of $\omega\sim E_T$ at low electron density we may  expand the square root in Eq.~\eqref{TStart} to obtain
\begin{eqnarray}\label{TExpand1}
\Sigma_x(\mathbf{0},\omega)\approx \int_{|\vecq|<k_F} \frac{d^2q}{(2\pi)^2}  T_{xe}\Big[\omega+\epsilon_\vecq-\epsilon_\vecq^\text{Tot}-\epsilon_F^R +i 0^+\Big]. \nonumber\\
\end{eqnarray}

Inserting the expression for $T_{xe}$ given by Eq.~\eqref{T2B} yields
\begin{eqnarray}
\Sigma_x(\mathbf{0},\omega)\approx\frac{2 \pi}{m}\int_0^{k_F} \frac{dq q}{2\pi}\frac{1}{\ln\left[\frac{E_T}{\omega-\epsilon_F^R +\gamma \vecq^2 +i0^+}\right]+i \pi}\nonumber\\ \label{T4}
\end{eqnarray}
where we define $\gamma = \frac{1}{2m_e}-\frac{1}{2 m_{Tot}}$ with $m_{Tot} =m_e + m_x$.

Eq.~\eqref{T4} can be integrated analytically and gives
\begin{eqnarray}\label{TAna}
\Sigma_x(\mathbf{0},\omega)\approx\frac{E_T}{2 m \gamma }\Bigg[\text{li}\left(\frac{-\omega +\epsilon_F^R - \gamma k_F^2}{E_T}\right)-\text{li}\left(\frac{-\omega +\epsilon_F^R }{E_T}\right)\Bigg]\nonumber\\
\end{eqnarray}
where $\text{li}(x)$ is the logarithmic integral function. Expanding to second order in $k_F$ finally yields
\begin{equation}
    \Sigma_x(\mathbf{0},\omega)\approx -\frac{k_F^2}{2m}\frac{1}{\ln\left(\frac{-\omega}{E_T}\right)} = n_e T_{xe}(0,\omega)
\end{equation}
which proves  Eq.~\eqref{SimpSig}.



\bibliography{biblio}

\end{document}